\documentclass[prb, showpacs, preprintnumbers, amsmath, amssymb, superscriptaddress, twocolumn]{revtex4}
\usepackage{graphicx, color, dcolumn, bm, natbib, amssymb, amsmath}
\usepackage[final=true]{hyperref}

\newcommand{\m}{\mathbf}

\newcommand{\p}{\parallel}
\newcommand{\ve}{\varepsilon}

\renewcommand{\o}{\omega}
\renewcommand{\t}{\theta}
\newcommand{\be}{\begin{eqnarray}}
\newcommand{\ee}{\end{eqnarray}}
\newcommand{\nn}{\nonumber}
\newcommand{\re}{\color{red}}

\begin{document}

\title{Small-angle neutron scattering in fully polarized phase of non-collinear magnets with interfacial-like DMI}

\author{Oleg I. Utesov}
\affiliation{Saint Petersburg State University, Ulyanovskaya 1, St. Petersburg 198504, Russia}
\affiliation{Petersburg Nuclear Physics Institute, Gatchina 188300, Russia}
\affiliation{St. Petersburg School of Physics, Mathematics, and Computer Science, HSE University, St. Petersburg 190008, Russia}

\begin{abstract}

Spin waves in non-collinear magnets with $C_{nv}$ symmetry are discussed in the context of inelastic neutron scattering in small-angle scattering geometry. In the framework of a minimal model including exchange coupling, Dzayloshinskii-Moriya interaction (DMI), and single-ion anisotropy, we consider the system properties at moderate external magnetic fields and in the high-field fully polarized phase. In the latter case, the magnon spectrum is gapped and non-reciprocal due to DMI with the minimum perpendicular to the field direction for the in-plane field and is ferromagnetic-like for the field along the high-symmetry axis. Inelastic small-angle neutron scattering in the fully polarized phase (SWSANS) is considered in four different geometries. It is shown that analysis of the field dependence of the SWSANS cutoff feature in these geometries allows determining all important parameters of the model. The possibility of the proposed method utilization in thin films with interfacial DMI is also discussed.

\end{abstract}

\maketitle

\section{Introduction}

It is well-known that Dzyaloshinskii-Moriya interaction~\cite{dzyaloshinsky1958, moriya1960} (DMI) can lead to non-collinear magnetic structures~\cite{dzyaloshinsky1958, dzyaloshinskii1964}. Despite many years passed since the first observation of helical structures in non-centrosymmetric magnets~\cite{ludgren1970}, this type of compounds still attracts significant attention. In particular, it is stimulated by their omnifarious phase diagrams, which include regions hosting topologically nontrivial phases~\cite{bogdanov1989,bogdanov1994,muhlbauer2009}. Moreover, isolated skyrmions and their ordered arrays (skyrmion lattices, SkLs) have several promising technological applications~\cite{fert2013}. These magnetic structures can also be stabilized in layered nanostructures with interfacial DMI (iDMI)~\cite{fert1980, fert2017}.

In this context, the developing of characterization methods of skyrmion-hosting materials becomes more and more demanded by material science and magnetism communities. Recently, the inelastic small-angle neutron scattering (SANS) on spin waves~\cite{toperverg1983,okorokov1986} (called SWSANS below) was shown to be fruitful for various chiral cubic helimagnets (e.g., MnSi) studies in fully-polarized by external field phase~\cite{grig2015,siegfried2017,grig2018jmmm,grig2018prb,ukleev2022}. Using this method, important information about spin-wave dynamics can be obtained. The latter is based on the theoretical prediction by Kataoka~\cite{kataoka1987}. In simple words, the SWSANS intensity is distributed on a three-dimensional sphere, which results in a circular spot when projected on a detector plane in the experiment. The radius of the spot is related to spin-wave stiffness and energy gap in the spectrum, whereas the center corresponds to the spiral vector. Moreover, even magnon damping can be studied using this technique, see Refs.~\cite{deriglazov1992, ukleev2022}.

In another type of non-centrosymmetric systems, so-called polar magnets, DMI favors cycloidal magnetic structures and N\'{e}el skyrmions and SkLs, which were indeed observed in Refs.~\cite{kezsmarki2015neel,kurumaji2017}. The same is also true for thin films with iDMI. The symmetry of such systems is $C_{nv}$ which is lower than the cubic one. So, we can expect different inelastic SANS maps with conventional B20 helimagnets pictures.

In the present research, we show that the SWSANS technique can be also used for the characterization of the compounds with interfacial-like DMI. Combining different experimental geometries gives a possibility for a comprehensive description of the system properties, such as spin-wave stiffness constants for in- and out-of-plane directions, DMI constant, and single-ion anisotropy. These parameters are analytically connected with the SWSANS cutoff curves in different azimuthal directions (not usually circles, in contrast to the cubic helimagnets) in four various cases.

The rest of the paper is organized as follows. In Sec.~\ref{SecModel} we present the model under investigation. Its properties are briefly summarized in Sec.~\ref{SecLow}. Sec.~\ref{SecFP} addresses the magnon spectrum in the fully polarized by external field phase for the two cases of in-plane and perpendicular field. Small-angle neutron scattering in the fully polarized phase is discussed in Sec.~\ref{SecSANS}, where four particular geometries of the experiment are proposed and analyzed theoretically. Sec.~\ref{SecConc} contains a discussion related to thin films and our conclusions.

\section{Model}
\label{SecModel}

\begin{figure}
  \centering
  \includegraphics[width=4cm]{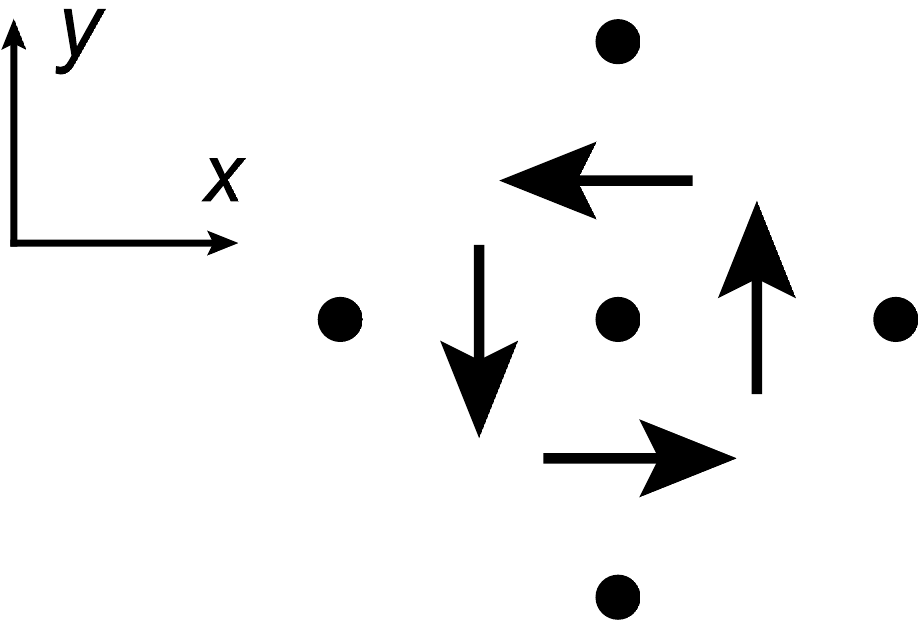}\\
  \caption{Scheme of the Dzyaloshinskii-Moriya interaction considered in the present study. DMI is allowed in the $xy$ plane perpendicular to the high-symmetry $z$ axis. Vectors of DMI interactions with nearest neighbors for the central magnetic ion are shown by the arrows. In the thin film case, the $xy$ plane is parallel to interfaces, which breaks the inversion symmetry. }\label{figDMI}
\end{figure}

We consider a spin Hamiltonian, which includes exchange coupling, Dzyaloshinskii-Moriya interaction, single-ion anisotropy, and Zeeman energy:
\be \label{ham1}
  \mathcal{H} &=& \mathcal{H}_{\textrm{EX}} + \mathcal{H}_{\textrm{DMI}} + \mathcal{H}_{\textrm{AN}} + \mathcal{H}_\textrm{Z}, \nn \\
  \mathcal{H}_{\textrm{EX}} &=& -\frac{1}{2} \sum_{\m{R},\m{R}^\prime} J_{\m{R}-\m{R}^\prime} \m{S}_\m{R} \cdot \m{S}_{\m{R}^\prime}, \nn \\
  \mathcal{H}_{\textrm{DMI}} &=& \frac12 \sum_{\m{R},\m{R}^\prime} \m{D}_{\m{R} - \m{R}^\prime} \cdot \left[ \m{S}_\m{R} \times \m{S}_{\m{R}^\prime} \right], \\
  \mathcal{H}_{\textrm{AN}} &=& - K \sum_{\m{R}} \left(S^z_\m{R}\right)^2, \nn \\
  \mathcal{H}_\textrm{Z} &=& - \m{h} \cdot \left( \sum_\m{R} \m{S}_\m{R} \right). \nn
\ee
Here $\m{R}$ enumerates all spins, which are arranged for definiteness in a simple tetragonal lattice with $C_{4v}$ symmetry. We choose the $z$ axis to be a high-symmetry direction. According to $\mathcal{H}_{\textrm{AN}}$, it is the easy axis for $K>0$ and the hard one for $K<0$ (the easy plane case). In Zeeman term, we use standard description of the magnetic field in energy units, so $\m{h} = - g \mu_B \m{H}$. DMI in the present system deserves more detailed discussion. In general, $C_{nv}$ symmetry allows for the following Lifshitz invariant in energy density~\cite{bogdanov1989} ($\m{M}$ is the magnetization):
\be
  W = \gamma \left[ M_z \frac{\partial M_x}{\partial x} - M_x \frac{\partial M_z}{\partial x} + M_z \frac{\partial M_y}{\partial y} - M_y \frac{\partial M_z}{\partial y}\right]
\ee
Microscopically, this form can be reproduced by the scheme shown in Fig.~\ref{figDMI}. Note, that this type of DMI is relevant to interfacial DMI, which arises due to symmetry breaking on the boundary between the magnetic material and another material~\cite{fert1980} (usually, heavy metal with large spin-orbit coupling). We also would like to point out that in our model we neglect magnetodipolar interaction, which can be important in multilayered systems with iDMI and can lead to a hybrid Bloch/N\'{e}el-type helicoids~\cite{legrand2018}. However, we focus on the effect of the iDMI on SWSANS spectra, where the dipolar forces are expected to play a minor role.

The subsequent discussion is much easier in the reciprocal space, so we introduce the Fourier transform on the lattice
\be \label{fourier1}
  \m{S}_\m{R} = \frac{1}{\sqrt{N}} \sum_{\m{q}} \m{S}_\m{q} e^{i \m{q} \cdot \m{R}},
\ee
where $N$ is the number of lattice sites. Then, the counterparts of various interactions in the Hamiltonian~\eqref{ham1} read
\be \label{ham2}
  \mathcal{H}_{\textrm{EX}} &=& - \frac12 \sum_{\m{q}} J_{\m{q}} \m{S}_\m{q} \cdot \m{S}_{-\m{q}}, \nn \\
  \mathcal{H}_{\textrm{DMI}} &=& \frac12 \sum_{\m{q}} \m{D}_{\m{q}} \cdot [\m{S}_\m{q} \times \m{S}_{-\m{q}}], \nn \\
  \mathcal{H}_{\textrm{AN}} &=&  - K \sum_{\m{q}} S^z_\m{q} S^z_{-\m{q}}, \\
  \mathcal{H}_{\textrm{Z}} &=& -\sqrt{N} \m{h} \cdot \m{S}_\m{0}. \nn
\ee
Below we assume, that the exchange is ferromagnetic and for small $q$ it can be expanded as follows:
\be \label{stiff}
  J_\m{q} \approx J_\m{0} - \frac{A_\perp (q^2_x + q^2_y)}{S} - \frac{A_\p q^2_z}{S}.
\ee
Here $A_\perp$ and $A_\p$ are corresponding spin-wave stiffnesses. Note also that in long-wavelength limit symmetry-allowed anisotropic exchange terms with sufficient accuracy can be mimicked by the single-ion anisotropy. However, the anisotropic exchange can be responsible for some fine effects, e.g., the modulation vector modulus dependence on the spin structure orientation. It was directly shown in FeGe in Ref.~\cite{ukleev2021}.

The fourier transform of DMI in the nearest-neighbors approximation is given by
\be
  \m{D}_\m{q} = - i D \sum_{b} \sin{\m{q} \cdot \m{b}} [\hat{z} \times \m{b}] \approx - i D \sum_{b} \m{q} \cdot \m{b} [\hat{z} \times \m{b}],
  \label{dq1}
\ee
where $b$ stands for all the bonds with nearest-neighbors of a certain spin, see Fig.~\ref{figDMI}. This formula can be further simplified using the property ($\alpha$ and $\beta$ vector components $x,y,z$)
\be
  \sum_b b_\alpha b_\beta = 2 \delta_{\alpha\beta}.
\ee
So, in the case of a tetragonal lattice, we have Eq.~\eqref{dq1} in the form
\be \label{dq2}
  \m{D}_\m{q} = - 2 i D [\hat{z} \times \m{q}].
\ee
Note that for hexagonal lattice it also holds but with a factor $3$ instead of $2$, so the discussion above can be easily applied in this case too. Evidently, the Fourier transform of DMI is lying in the $xy$ plane and is perpendicular to the corresponding momentum.

\section{Moderate fields}
\label{SecLow}

We start our analysis of the above model from the case of relatively small fields, where at $T=0$ non-collinear spin structures (cycloids or N\'{e}el skyrmions) are possible. Here we neglect the single-ion anisotropy term in Hamiltonian~\eqref{ham1} (it can provide higher harmonics or even destroy non-collinear structures~\cite{izyumov1983neutron} if it is strong enough). For description of single-modulated structures we use the Kaplan helix representation~\cite{Kaplan1961}:
\be
  \label{kaplan}
  \m{S}_\m{R} = S \left( \m{A} e^{i \m{k} \cdot \m{R}} + \m{A}^* e^{- i \m{k} \cdot \m{R}} \right) \cos{\alpha} + S \hat{c} \sin{\alpha}.
\ee
Here $\alpha$ is the cone angle, $\m{A} = (\hat{a} - i \hat{b})/2, \, \m{A}^* = (\hat{a} + i \hat{b})/2$, $\hat{a},\hat{b},\hat{c}$ is some orthogonal basis. Well-known particular cases are a screw spiral with modulation vector $\m{k} \parallel \hat{c}$ and a cycloid with $\m{k} \perp \hat{c}$. Useful properties for calculations are the following: $\m{A}^2=0, \, \m{A} \cdot \m{A}^* = 1/2, \, \hat{c} \times \m{A} = i \m{A}, \, \m{A} \times \m{A}^* = i \hat{c}/2$. From Eq.~\eqref{kaplan}, one can deduce
\be
  \m{S}_\m{k} &=& \sqrt{N} S \m{A} \cos{\alpha} , \nn \\ \m{S}_{-\m{k}} &=& \sqrt{N} S  \m{A}^* \cos{\alpha}, \\ \m{S}_\m{0} &=& \sqrt{N} S \hat{c} \sin{\alpha}. \nn
\ee
All the harmonics with other $\m{q}$ are zero. Plugging these formulas into Hamiltonian~\eqref{ham2}, one can calculate the spin structure energy per one spin
\be \label{ekaplan1}
  E &=& - \frac{S^2}{2} \left( J_\m{k} \cos^2{\alpha} + J_\m{0} \sin^2{\alpha} \right) - \frac{i S^2}{2} \m{D}_\m{k} \cdot \hat{c} \cos^2{\alpha}  \nn \\ &&- S \m{h}\cdot \hat{c} \sin{\alpha}.
\ee
For small $k$ we can rewrite it using Eqs.~\eqref{stiff} and~\eqref{dq2} as follows:
\be \label{ekaplan2}
  E &=& - \frac{S^2}{2} J_\m{0} + \frac{S (A_\perp k^2_\perp + A_\p k^2_\parallel) \cos^2{\alpha}}{2}  \nn \\
  && - S^2 D \hat{c} \cdot [\m{k} \times \hat{z}] \cos^2{\alpha} - S \m{h}\cdot \hat{c} \sin{\alpha}.
\ee
Here we divide the modulation vector into two parts, $k_\parallel$ is along the $z$ axis and $k_\perp$ lies in the $xy$ plane. Evidently, minimal $E$ requires $k_\parallel = 0$. For in-plane magnetic fields we have $\hat{c}$ along $\m{h}$ and $\m{k}$ along $D \hat{z} \times \m{h}$ (its direction is dependent on the sign of $D$). So, in general, the solution is the conical cycloid, which energy~\eqref{ekaplan2} can be rewritten as
\be \label{ekaplan3}
  E &=& - \frac{S^2}{2} J_\m{0} + \frac{S \cos^2{\alpha}}{2}(A k^2_\perp - 2 D k_\perp) - S h \sin{\alpha}. \nn \\
\ee
Its minimization with respect to $\alpha$ and $k_\perp$ yields
\be \label{sp1}
  k_\perp &=& \frac{S D}{A_\perp} \equiv k, \\ \label{alpha1}
  \sin{\alpha} &=& \frac{h}{h_{C2}}, \, h \leq h_{C2}, \\ \label{hc21}
  h_{C2} &=& A_\perp k^2.
\ee
At fields $h \geq h_{C2}$ the system is in the fully polarized phase. Notation $C2$ is related to the fact, that at very small fields in-plane anisotropy and higher order in $\m{q}$ terms in Eq.~\eqref{dq1} play important role in the cycloid orientation determination~\cite{maleyev2019}, and the regime $\hat{c} \parallel \m{h}$ is correct for moderate fields \mbox{$h > h_{C1}$} only. The field $h_{C1}$ accurate description requires accounting for symmetry-allowed in-plane anisotropy (e.g., quadratic or hexagonal) and the anisotropic exchange, which deserves a separate study.

In the case of iDMI (e.g., a thin film of a ferromagnet with neighboring nonmagnetic materials), the equations above should be modified. Let us consider a system with $M$ layers of magnetic material and two interfaces with DMI constants $D_1$ and $D_2$. Due to the opposite directions of perpendicular to interfaces vectors, the effective DMI reads
\be
  D_{eff} = \frac{D_1-D_2}{M}.
\ee
which should be used in the energy function~\eqref{ekaplan1}.

All the layers feel the exchange stiffness and the external field, but the DMI is only on the interfaces. Then, the parameters of the cycloid solution alter according to the following formulas:
\be \label{sp2}
  k &=& \frac{S D_{eff}}{A_\perp}, \\
  \sin{\alpha} &=& \frac{h}{h_{C2}}, \, h < h_{C2}, \\
  h_{C2} &=& A_\perp k^2 = \frac{S^2 D^2_{eff}}{A_\perp}.
\ee
We see, that they are identical to the bulk ones~\eqref{sp1},~\eqref{alpha1}, and~\eqref{hc21} upon the substitution $D \rightarrow D_{eff}$.

\section{Fully polarized phase}
\label{SecFP}

\subsection{In-plane field}

Here we consider the spin-wave spectrum in a relatively large in-plane external field $h \geq h_{C2}$. We neglect small in-plane anisotropic terms, so the result is independent of the magnetic field orientation in the $xy$ plane. We choose $x$ axis along $\m{h}$ and use the following approximate Holstein-Primakoff spin operators representation~\cite{Holstein1940}:
\be \label{spinrep}
  S^x_\m{R} &=& S - a^\dagger_\m{R} a_\m{R}, \nn \\
  S^y_\m{R} &=& \sqrt{\frac{S}{2}} \left(a^\dagger_\m{R} + a_\m{R}\right), \\
  S^z_\m{R} &=& i \sqrt{\frac{S}{2}} \left(a^\dagger_\m{R} - a_\m{R}\right).
\ee
The Fourier transform of magnon creation-annihilation operators reads [cf. Eq.~\eqref{fourier1}]
\be \label{fourier2}
  a_\m{R} = \frac{1}{\sqrt{N}} \sum_{\m{q}} a_\m{q} e^{i \m{q} \cdot \m{R}}, \quad
  a^\dagger_\m{R} = \frac{1}{\sqrt{N}} \sum_{\m{q}} a^\dagger_\m{q} e^{-i \m{q} \cdot \m{R}}.
\ee
Using these formulas, we obtain spin operators components in reciprocal space in the form
\be \label{spinrep2}
  S^x_\m{q} &=& \sqrt{N} S \delta_{\m{q},\m{0}} - \frac{1}{\sqrt{N}} \sum_\m{p} a^\dagger_\m{p} a_{\m{p} + \m{q}}, \nn \\
  S^y_\m{q} &=& \sqrt{\frac{S}{2}} \left(a^\dagger_{-\m{q}} + a_\m{q}\right), \\
  S^z_\m{q} &=& i \sqrt{\frac{S}{2}} \left(a^\dagger_{-\m{q}}- a_\m{q}\right). \nn
\ee

Next, one can calculate bilinear in Bose-operators part of Hamiltonian~\eqref{ham2}, which reads
\be \label{hamb1}
  \mathcal{H}^{(2)}_{\textrm{EX}} &=& S \sum_{\m{q}} \left( J_\m{0} - J_\m{q} \right) a^\dagger_\m{q} a_{\m{q}}, \nn \\
  \mathcal{H}^{(2)}_{\textrm{DMI}} &=& - 2 S D\sum_{\m{q}} q_y a^\dagger_\m{q} a_{\m{q}}, \\
  \mathcal{H}^{(2)}_{\textrm{AN}} &=& - S K \sum_{\m{q}} \left( a^\dagger_\m{q} a_{\m{q}} + \frac{a_{\m{q}} a_{-\m{q}} + a^\dagger_{\m{q}} a^\dagger_{-\m{q}}}{2} \right), \nn \\
  \mathcal{H}^{(2)}_{\textrm{Z}} &=& h \sum_{\m{q}} a^\dagger_\m{q} a_{\m{q}}. \nn
\ee
So, without contribution from the anisotropy, the result is very simple:
\be \label{hamb2}
  \mathcal{H}^{(2)} &=& \sum_\m{q} \left[ S\left( J_\m{0} - J_\m{q} \right) - 2 S D q_y + h   \right] a^\dagger_\m{q} a_{\m{q}} \nn \\ &\equiv& \sum_\m{q} \varepsilon_\m{q} a^\dagger_\m{q} a_{\m{q}},
\ee
where $\varepsilon_\m{q}$ is the magnon energy. For small $q$ it can be further simplified,
\be \label{specin}
  \varepsilon_\m{q} &=& A_\perp \left[q^2_x + (q_y - k)^2 \right] + A_\p q^2_z + h - h_{C2} \Longleftrightarrow \nn \\
   \varepsilon_\m{q} &=& A_\perp (\m{q}_\perp - \m{k})^2 + A_\p q^2_z + \Delta.
\ee
This result can be compared with the one by Kataoka for B20 helimagnets~\cite{kataoka1987}. The similarity is that the gap $\Delta = h - h_{C2}$ and the spectrum is non-reciprocal  ($\varepsilon_\m{q} \neq \varepsilon_{-\m{q}}$) due to DMI. However, the spectrum minimum lies perpendicular to $\m{h}$ direction (along $\m{k}$ -- the cycloid modulation vector).


\subsection{The role of magnetic anisotropy in spin wave spectrum}

Taking into account $\mathcal{H}^{(2)}_{\textrm{AN}}$ term in Eq.~\eqref{hamb1}, we obtain the bilinear part of the Hamiltonian for small $q$ in the following form [cf. Eq.~\eqref{hamb2}]:
\be
  \mathcal{H}^{(2)} &=& \sum_\m{q} \left( E_\m{q} a^\dagger_\m{q} a_{\m{q}} + B_\m{q} \frac{a_{\m{q}} a_{-\m{q}} + a^\dagger_{\m{q}} a^\dagger_{-\m{q}}}{2} \right), \nn \\
  E_\m{q} &=&  A_\perp \left[q^2_x + (q_y - k)^2 \right] + A_\p q^2_z + h - h_{C2} - S K , \nn \\
  B_\m{q} &=& - S K.
\ee
Note that $E_\m{q} \neq E_{-\m{q}}$. Then, the generalized Bogoliubov transformation should be used to obtain the magnon spectrum:
\be \label{specinan}
  \varepsilon^\prime_\m{q} &=& \frac{E_\m{q} - E_{-\m{q}}}{2} + \sqrt{\left(\frac{E_\m{q} + E_{-\m{q}}}{2} \right)^2 - B^2_\m{q}}.
\ee
First, we see, that the anisotropy renormalizes the critical field, which in the limit of $S |K| \ll A_\perp k^2$ reads $h^\prime_{C2} = A_\perp k^2 + S K$. Second, for $\varepsilon_\m{q} \gg S |K|$ ($\m{q}$ not very close to $\m{k}$ or magnetic field not very close to $h_{C2}$) we obtain
\be \label{specinan2}
  \varepsilon^\prime_\m{q} &\approx& \varepsilon_\m{q} -S K,
\ee
which can be considered as an effective gap renormalization $\Delta \rightarrow \Delta - S K$.

It is pertinent to note that similar effects can be a consequence of the magnetodipolar interaction. Moreover, strong (in the units of the characteristic cycloid energy $A_\perp k^2$) anisotropy or dipolar forces can significantly change the spectrum making it linear in some range of $\m{q}$. In order to use the SWSANS technique discussed in Sec.~\ref{SecSANS}, spin-wave energy near the cutoff momenta should be much larger than the anisotropic interaction energies [in this case, the expansion~\eqref{specinan2} of Eq.~\eqref{specinan} is correct].


\subsection{Perpendicular field $\m{h} \p \hat{z}$}

It is also useful to consider the spectrum of the fully polarized phase in the magnetic field along $\hat{z}$. In this case, in spin operators quantization rules~\eqref{spinrep2} one should make the following substitutions $x \rightarrow z, \, y \rightarrow x, \, z \rightarrow y $. It is easy to see that after these substitutions, DMI includes only terms with an odd number of Bose-operators, because cross product always includes $S^z$ component [see Eqs.~\eqref{ham2},~\eqref{dq2}, and~\eqref{spinrep2}]. Moreover, linear terms vanish because $\m{D}_\m{0}=\m{0}$. So, in the linear spin-wave theory DMI does not influence the magnon spectrum. However, in contrast with usual ferromagnets, there will be some quantum corrections to the spectrum even at $T=0$ due to DMI.

Next, the contribution from the single-ion anisotropy reads
\be
  \mathcal{H}^{(2)}_\textrm{AN} =  2 S K \sum_\m{q} a^\dagger_\m{q} a_{\m{q}},
\ee
which along with terms from exchange coupling and Zeeman term [see Eq.~\eqref{hamb1}] gives the magnon spectrum
\be \label{specperp}
  \varepsilon_\m{q} &=& A_\perp q^2_\perp + A_\p q^2_z + h + 2 S K \nn \\ &\equiv&  A_\perp q^2_\perp + A_\p q^2_z+\Delta^\prime.
\ee
It indicates that even at $h=0$ perpendicular collinear phase can be observed as metastable for the easy axis anisotropy, whereas in the easy plane case one needs $h \geq 2 S |K|$. We denote the gap here as $\Delta^\prime$ to avoid confusion with the different one for the in-plane field ($\Delta$), see Eq.~\eqref{specin}.

To conclude this section, we would like to point out that in the case of structures with iDMI, the Fourier transform along the $z$ axis can become incorrect. For instance, in a single thin film of magnetic material, $A_\p$ and $q_z$ become meaningless quantities. So, in Eqs.~\eqref{specin},~\eqref{specinan2} and~\eqref{specperp} the contribution $A_\p q^2_z$ should be omitted. Instead, due to the size quantization effect, some correction to $\Delta$ can be expected, which should essentially depend on the boundary conditions.

\section{Small-angle neutron scattering in fully polarized phase}
\label{SecSANS}

Here we discuss how inelastic small-angle neutron scattering on magnons can be used to obtain the corresponding material parameters from the experiment. In particular, we derive expressions for the so-called cutoff angle in four various experimental geometries. Note that the equations below are written for unpolarized neutrons; however, generalization to the polarized case is straightforward~\cite{maleev2002}. The latter can be important to pinpoint magnetic scattering~\cite{toperverg1983,okorokov1986}. We also would like to point out that the sign of $D$ can be only determined with the polarized neutrons~\cite{maleyev1995,grigoriev2009}.

So, in our case, we use the following equation for neutron scattering cross-section (see, e.g., Ref.~\cite{maleev2002}):
\be \label{sigma1}
  \sigma(\o,\m{Q}) &=& \frac{1}{\pi} \frac{k_f}{k_i} \left[ 1 - \exp{\left( -\frac{\o}{T} \right)}\right]^{-1}  r^2 |F_m|^2 \nn \\ && \mathrm{Im} \, \chi^{(S)}_{\alpha \beta}(\o,\m{Q}) (\delta_{\alpha\beta} - \hat{Q}_\alpha \hat{Q}_\beta).
\ee
Here $k_i(k_f)$ is incident (scattered) neutron momentum, $\o$ and $\m{Q}$ are transferred energy and momentum, respectively, $r$ is the classical electron radius, $F_m$ is the magnetic form factor of the ions and $\chi^{(S)}$ is the symmetric part of magnetic susceptibility. The latter in our case reads (diagonal components of the transverse susceptibility)
\be
  \mathrm{Im}\,\chi^\perp(\o,\m{Q})  = \frac{\pi \langle S \rangle}{2} \left[ \delta(\o-\ve_\m{Q}) -  \delta(\o+\ve_{-\m{Q}}) \right],
\ee
where the terms in brackets correspond to emission and absorbtion of magnons and $\langle S \rangle$ is thermodynamical average of the spin.

Next, as in Ref.~\cite{grig2015}, we consider the case $\o \ll T$ and replace $\left[ 1 - \exp{\left( -\o/T \right)}\right]^{-1}$ by $T/\o$ in Eq.~\eqref{sigma1}. Moreover, $k_f \approx k_i$ because the momentum transfer is small. Finally, in the SWSANS experiments, one should average the cross-section over $\omega$, so we obtain
\be \label{sigma2}
  \sigma(\m{Q}) &\propto& \langle S \rangle T \int \frac{d \o}{\o} \left( 1 + \frac{\hat{\m{Q}} \cdot \m{h}}{h} \right) \\ \nn &&\left[ \delta(\o-\ve_\m{Q}) -  \delta(\o+\ve_{-\m{Q}}) \right].
\ee
Here we explicitly use the fact that the sample is magnetized along the field. Particular scattering and magnetic field geometries are considered separately below. In the obtained results, three dimensionless parameters are important:
\be \label{Aanis}
  a &=& \frac{A_\p}{A_\perp}, \\
  \t_0 &=& \frac{E_i}{A_\p k^2_i}, \\
  \t_B &=& \frac{k}{k_i}.
\ee
The first one measures the stiffness anisotropy, the second is the ratio of the incident neutron energy to the certain characteristic magnetic energy and the third one is the so-called Bragg angle, indicating the position of the scattering peak from an incommensurate order in the detector plane.

\begin{figure}
  \centering
  \includegraphics[width=4.2cm]{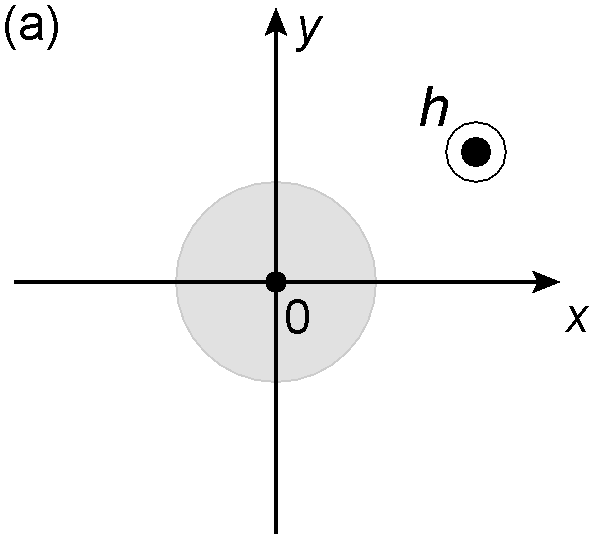}
  \hfill
  \includegraphics[width=4.2cm]{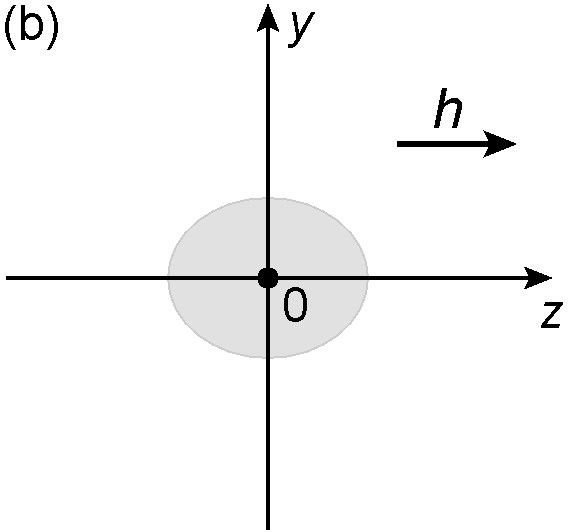}
  \vspace{1cm}
  \includegraphics[width=4.2cm]{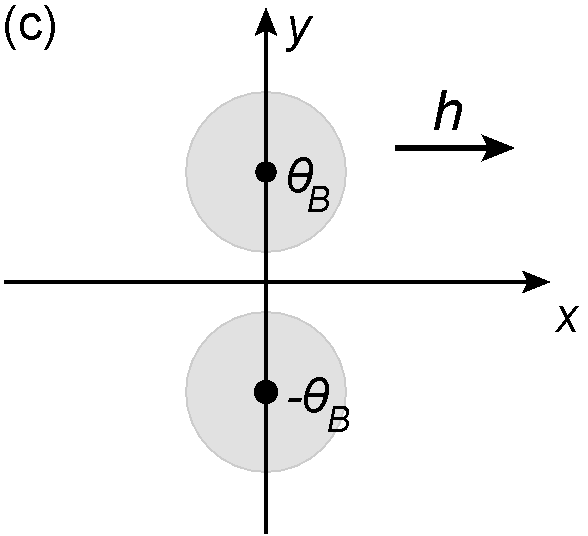}
  \hfill
  \includegraphics[width=4.2cm]{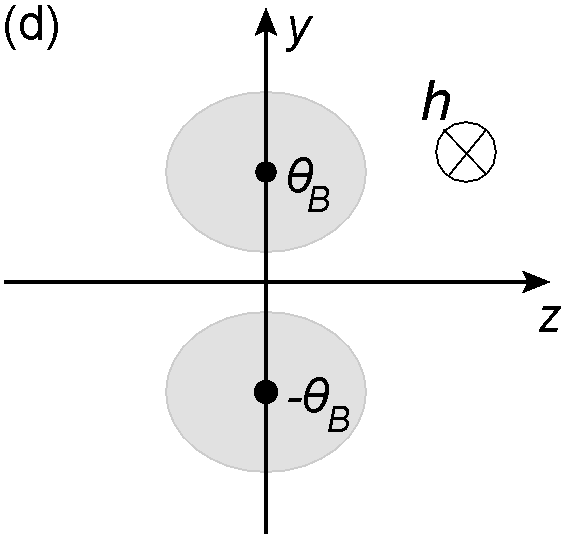}
  \caption{Sketches of SWSANS maps in four different geometries. Stiffness anisotropy parameter $a=0.64$ [see Eq.~\eqref{Aanis}] is used. In the case of $\mathbf{H} \p \hat{z}$ [(a) and (b), perpendicular field] the magnon spectrum is reciprocal and the scattering is ferromagnetic-like. Moreover, in panel (b) ellipticity of the cutoff angle related to $a$  is pronounced. When $\mathbf{H} \p \hat{x}$ (in-plane field) the magnon spectrum is non-reciprocal, and the SANS signals are centered at the Bragg angles [(c) and (d)]. For the scattering in the $yz$ plane, the ellipticity is also visible. In the case of polarized neutrons, in panels (c) and (d) contributions centered at $\pm \t_B$ have unequal intensities; the difference being $\propto Q^2_x/\m{Q}^2$, which can be used to determine the sign of $D$. }\label{figmaps}
\end{figure}

The findings of the subsections below are summarized in Fig.~\eqref{figmaps}. Here it is pertinent to note that in cubic helimagnets with DMI of the bulk type, one cannot observe ferromagnetic-like scattering patterns shown in Figs.~\eqref{figmaps}(a) and (b). Moreover, the corresponding SWSANS maps are always circles due to the cubic symmetry, which are shifted along magnetic field~\cite{grig2015}, not perpendicular to the field direction as for systems with interfacial-like DMI [see Figs.~\eqref{figmaps}(c) and (d)].

\subsection{Scattering in $xy$ plane, $\m{h} \p \hat{z}$}

In this case, the transferred momentum can be written in the following way:
\be \label{Q1}
  \m{Q} = k_i \left(\t_x,\t_y,\frac{\o}{2 E_i} \right),
\ee
where $E_i$ is the energy of the incident neutron. Under conditions of SANS experiment, $Q \ll k_i$ and $\t_x, \t_y$ measure a point in a detector (scattering ``angles'').

Next, delta-functions in Eq.~\eqref{sigma2} determine the so-called cutoff in the detector plane at which the cross-section diverges (in an integrable way) and after which the signal is zero (however, it lasts beyond the cutoff due to finite magnon lifetime, see Ref.~\cite{ukleev2022} for the corresponding theory). The cutoff can be obtained by considering solutions of equation $\o = \ve_{\m{Q}}$ with respect to $\o$ (equation $\o + \ve_{-\m{Q}} =0$ yields the same physics). It is convenient to rewrite this equation using the variable $t = \o/2 E_i$, which after some transformations [we use the spectrum~\eqref{specperp}] reads
\be
  t^2 - 2 \t_0 t + \frac{\Delta^\prime}{A_\p k^2_i} + \frac{\t^2_x + \t^2_y}{a}=0.
\ee
So, in this case, the standard ferromagnetic-like picture arises: the scattering is located in a circle centered at $(\t_x,\t_y)=0$ and bounded by the cutoff satisfying the condition:
\be
   \frac{\t^2_C}{a} = \t^2_0 - \frac{\Delta^\prime}{A_\p k^2_i}.
\ee
Under the magnetic field growth{\re ,} $\t^2_C$ linearly decreases (general property for all geometries considered in the present study), which can be used for the experimental data interpretation.

In Eq.~\eqref{sigma2} the following substitution should be done
\be
  1 + \frac{\hat{\m{Q}} \cdot \m{h}}{h} \rightarrow 1+ \frac{(\o/2E_i)^2 }{\t^2_x + \t^2_y  + (\o/2E_i)^2 }.
\ee
Importantly, one can see that it is isotropic in the $\t_x \t_y$ plane.

\subsection{Scattering in $xz$ plane, $\m{h} \p \hat{z}$}

In this case, we have [cf. Eq.~\eqref{Q1}]
\be \label{Q2}
  \m{Q} = k_i \left(\t_x, \frac{\o}{2 E_i}, \t_z \right),
\ee
Equation $\o = \ve_{\m{Q}}$ is equivalent to
\be
  t^2 - 2 a \t_0 t + \frac{\Delta^\prime}{A_\perp k^2_i} + \t^2_x + a \t^2_z=0.
\ee
So, the cutoff line in the $\t_x\t_z$ plane here is an ellipse. Explicitly:
\be
  \t^2_x + a \t^2_z = a^2 \t^2_0 - \frac{\Delta^\prime}{A_\perp k^2_i}.
\ee
Importantly, the ratio of this ellipse semi{\re -}axes can be directly used for the parameter $a$ determination. Furthermore, the cross-section~\eqref{sigma2} acquires weak angular dependence,
\be
   1 + \frac{\hat{\m{Q}} \cdot \m{h}}{h} \rightarrow 1+ \frac{\t^2_z}{\t^2_x + (\o/2E_i)^2 + \t^2_z }.
\ee

\subsection{Scattering in $xy$ plane, $\m{h} \p \hat{x}$}

In this case, the magnon spectrum becomes non-reciprocal [see Eq.~\eqref{specin}]. So, the scattering patterns from $\o = \ve_{\m{Q}}$ and from $\o + \ve_{-\m{Q}} =0$ are centered in $(0, \t_B)$ and $(0, -\t_B)$, respectively, for magnetic field along $x$ axis. The result is a superposition of these two contributions, which can overlap. For brevity, below we consider the contribution from $\o = \ve_{\m{Q}}$ only; its counterpart can be obtained in a straightforward way.

In this geometry we have
\be \label{Q3}
  \m{Q} = k_i \left(\t_x, \t_y, \frac{\o}{2 E_i} \right),
\ee
however, the spectrum is given by Eq.~\eqref{specin} or by Eq.~\eqref{specinan2} if the single-ion anisotropy contribution is taken into account. It is convenient to use
\be
  \t_{rel} = \sqrt{\t^2_x + (\t_y - \t_B)^2},
\ee
which is the distance with respect to the Bragg angle in the detector plane. After some calculation we have the following equation for $t$:
\be
   t^2 - 2 \t_0 t + \frac{\Delta}{A_\p k^2_i} + \frac{\t^2_{rel}}{a}=0.
\ee
Hence, the cutoff is given by
\be
  \frac{\t^2_{relC}}{a} = \t^2_0 - \frac{\Delta}{A_\p k^2_i}.
\ee
In the cross-section~\eqref{sigma2} angle-dependent factor emerges from $1 + \hat{\m{Q}} \cdot \m{h}/h $ under the integration:
 \be
    1+ \frac{\t^2_x}{\t^2_x + \t^2_y + (\o/2E_i)^2  }.
\ee

\subsection{Scattering in $yz$ plane, $\m{h} \p \hat{x}$}
\label{SANSd}

Here the transferred momentum is
\be \label{Q4}
  \m{Q} = k_i \left(\frac{\o}{2 E_i}, \t_y,  \t_z \right).
\ee
Energy conservation law leads to
\be
   t^2 - 2 a \t_0 t + \frac{\Delta}{A_\perp k^2_i} + (\t_y - \t_B)^2 + a \t^2_z=0.
\ee
So, the cutoff line is the ellipse centered in $(0,\t_B)$ and satisfying the following equation:
\be
  (\t_y - \t_B)^2 + a \t^2_z = a^2 \t^2_0 - \frac{\Delta}{A_\perp k^2_i}.
\ee
In this case, in Eq.~\eqref{sigma2} the following substitution is in order
\be
  1 + \frac{\hat{\m{Q}} \cdot \m{h}}{h} \rightarrow 1+ \frac{(\o/2E_i)^2 }{\t^2_y + \t^2_z  + (\o/2E_i)^2 }.
\ee

\subsection{Connection with real compounds}

As an example for some estimations, we take VOSe$_2$O$_5$ compound, where experimentally $H_{C2} \approx 100$~Oe (thus, $h_{C2} \approx 0.001$~meV) and \mbox{$k \approx 0.046$~nm$^{-1}$} (see Ref.~\cite{kurumaji2017}). Assuming utilization of neutrons with $\lambda = 10~\textrm{\AA}$ (their energy is $\approx 0.8$~meV) and using the equations above, one can obtain the following parameters:
\be
  \t_B \approx 0.007, \quad \t_0 \approx 0.042.
\ee
So, when $H$ is close to $H_{C2}$ the contributions from negative and positive frequencies $\omega$ will overlap [in contrast to sketches shown in Fig.~\ref{figmaps}(c) and (d)], however the Bragg angle and the cutoff curves should be resolvable. Note that the magnon energy at the cutoff is essentially larger than rather small energy $h_{C2}$ (approximately by two orders of magnitude) because $\t_0 \gg \t_B$, which justifies utilization of the spectrum derived in Sec.~\ref{SecFP}.

In the thin film case, in order to make some estimations, we take parameters of the system modeled in Ref.~\cite{moon2013}. The modulation vector \mbox{$k \approx 0.04$~nm$^{-1}$} and the characteristic cycloid energy $A_\perp k^2 \approx 0.007$~meV. So, one has
\be
  \t_B \approx 0.006, \quad \t_0 \approx 0.0045.
\ee
We conclude that for these parameters scattering maps should look like the one shown in Fig.~\ref{figmaps}(c) (in that particular scattering geometry) with well-separated contributions.

Finally, we would like to point out that if the dipolar interaction or single-ion anisotropy is significant for relevant magnon momenta (near the cutoff) the developed in the present paper approach is inapplicable or applicable only on a semi-quantitative level [see also text after Eq.~\eqref{specinan2}].

\section{Discussion and conclusions}
\label{SecConc}

For thin film characterization{\re ,} various methods are usually used (see Ref.~\cite{kuepferling2020} and references therein). As it follows from the theoretical considerations, their properties in the fully polarized phase should be almost the same with bulk systems with the difference that the effective DMI should be used. However, the problem of weak signals in SWSANS measurements is expected to be crucial. Presumably, it can be overcome by stacking these layers with metallic spacers to increase scattering volume or using some off-specular scattering in reflective geometry (see, e.g., Refs.~\cite{lauter2006,zabel2007,toperverg2015}).

Another obstacle in the thin film case can be a possible breakdown of the parallel to high symmetry axis spin-wave stiffness $A_\p$ and the corresponding momentum $q_z$ notation. For example, in a single film case{\re ,} magnon states size-quantization effect can become important. In this case, one would have (cf. Subsec.~\ref{SANSd}) transferred momentum in the form of Eq.~\eqref{Q4} and the following energy conservation law:
\be
   t^2 - 2 \t^\prime_0 t + \frac{\Delta}{A_\perp k^2_i} + (\t_y - \t_B)^2=0,
\ee
where $\t^\prime_0 = E_i/ A_\perp k^2_i$ and the gap $\Delta$ can possible acquire some additional contribution due to the size quantization effect. This equation determines the cutoff along the $\t_y$ axis in the detector plane
\be
  |\t_y - \t_B|\Bigr|_C = (\t^\prime_0)^2 - \frac{\Delta}{A_\perp k^2_i}.
\ee
At the same time, the signal $\t_z$-dependence is expected to have power-law decaying tails with the characteristic scale $\t_{zC} \sim 1/(k_i d)^2$, where $d$ is the magnetic film thickness.

To conclude, inelastic small-angle neutron scattering on magnons in the fully polarized phase (SWSANS) is proposed as a tool for the determination of various parameters of ferromagnets with interfacial-like DMI, namely, spin-wave stiffness along and perpendicular to the high-symmetry axis, constant in DMI, and single-ion anisotropy. The method relies on the cutoff feature of SWSANS maps. The latter is theoretically connected with the system parameters in four various experimental geometries. We show that elliptical scattering patterns can be observed and the ratio between the corresponding semi-axes yields the ratio between spin wave stiffnesses. Furthermore, for the in-plane field, scattering patterns are centered at the so-called Bragg angle, which is straightly related to DMI magnitude. Finally, variation of the external magnetic field allows quantifying all the parameters listed above as well as the single-ion anisotropy.

Important differences with conventional SWSANS maps in cubic helimagnets can be summarized as follows. (i) In polar magnets, three various types of scattering patterns are predicted, namely, ferromagnetic-like, circular and elliptical, whereas in cubic systems only circular ones can be observed. (ii) In-plane magnetic field shifts the signal in the perpendicular direction, not along the field as in cubic helimagnets. (iii) Out-of-plane field does not produce any shift at all, and the scattering is ferromagnetic-like.

\begin{acknowledgments}

The work is dedicated to the blessed memory of S.V.\ Maleev who inspired this research. We are grateful to V.\ Ukleev and S.V. Grigoriev for valuable discussions. The reported study was funded by the Russian Federation President Grant No. MK-1366.2021.1.2.

\end{acknowledgments}

\bibliography{TAFbib}

\end{document}